**Selective disruption of reach-related saccade timing following a middle-cerebral artery stroke**

Mahya Beheshti*, Todd E Hudson*, Rajvardhan Gadde, Glenn Alvarez Arias, Karis Huh, Robert L Sainburg, JR Rizzo

*Co-first authors

**Introduction**

Coordinated control of eye and hand movements is critical for nearly all goal-directed actions, underpinning tasks ranging from simple object manipulation to complex tool use. Typically, gaze is directed toward the target before or during the movement, allowing the brain to extract spatial and contextual information that corrects and guides reach, grasping and manipulation of an intended target. The timing of these gaze shifts is not incidental, but rather necessary to enable functional interaction with the environment[1,2]. These gaze shifts operate with time-sensitive precision even in dynamic environments, for which both the target and the context may be in motion[3].

At the neural level, temporal coupling between reach and saccade is mediated by cortical and subcortical networks that include the posterior parietal cortex, frontal eye fields, premotor cortex, cerebellum and superior colliculus, which interact to integrate sensory input, define motor goals, and temporally couple movement initiation across effectors[4,5]. Within this system, the initiation of a reach is typically coordinated with a saccade in a highly stereotyped temporal pattern. Muscle activation often begins first in the high-inertia arm to compensate for its longer movement duration, while the low-inertia oculomotor system is recruited shortly afterward. Despite being activated second, the eyes typically acquire the target 100–300 ms prior to the hand, forming a consistent temporal relationship thought to reflect shared control mechanisms that couple movement initiation[6-9].

Stroke disrupts these processes at several levels. Beyond single-effector motor, dexterity, or coordination impairments, stroke can degrade eye–hand integration, thereby providing a powerful model to examine how these processes can break down, particularly in relation to eye-hand coordination deficits[10-12]. Importantly, deficits in coordination may persist even when individual effectors (e.g., eye or hand) retain near-normal function[10]. This "decoupling" affects real-world activities, reducing task efficiency and safety[10-12]. Understanding such deficits in eye-hand coordination provides critical insight into both functional outcomes and rehabilitative strategies[12].

This study tests stroke patients and age-matched controls in a series of tasks to dissociate motor execution from inter-effector coupling. We assessed and compared eye and hand behavior during: (1) a natural dual-task condition, in which participants simultaneously performed coordinated reaches and saccades; (2) single-task conditions that isolated either saccadic or reaching movements; and (3) a segmented condition, where saccades and reaches were temporally separated by distinct cues. Comparing behavior across these tasks, we sought to dissociate impairments in motor execution from impairments in inter-effector coupling. We hypothesized that stroke participants would retain the ability to generate both reaches and saccades in structured single-effector tasks (look or reach) but would exhibit impaired temporal coupling between the two in naturalistic conditions (look and reach). Specifically, we predicted that while controls would show a stereotyped pattern of saccades preceding reach termination, stroke participants would generate saccade timing distributions that are uncoupled from reach timing. Importantly, we anticipated that this uncoupling could be rescued by imposing external structure through cueing, as supported by evidence that biofeedback may re-coordinate decoupled saccade-reach timing in stroke[11].

## Methods

### Participants

Six participants participated in this research study, comprising three participants in the control group 53-75 and three participants in the stroke group 62–70. The study protocol was reviewed and approved by the Institutional Review Board of New York University's School of Medicine. All participants provided written consent before participation.

All participants underwent comprehensive medical history reviews and neurological and musculoskeletal examinations, including range-of-motion assessments to determine study eligibility. Stroke participants met the following inclusion criteria: age 18 years or older, history of middle cerebral artery (MCA) stroke occurring at least one month prior to enrollment, ability to complete the Fugl-Myer Assessment (FMA) with scores of 50-57 for upper extremities, demonstration of full horizontal and vertical eye movements as verified the experimenter, capability to perform pointing task, and willingness to complete all clinical evaluations. Additionally, participants were required to provide informed consent and complete HIPAA certification.

Both stroke and control participants were excluded if they presented with cognitive impairment (Mini-Mental Status Exam score < 24) or significant ocular issues, including eye injury, extraocular muscle weakness, or visual field deficits. Visual function was assessed using the Beery-Buktenica Developmental Test of Visual-Motor Integration, Snellen chart, National Eye Institute Visual Functioning Questionnaire, and a 10-item neuro-ophthalmic supplement survey. Additional exclusion criteria included the presence of hemi-spatial neglect (assessed through line bisection and single-letter cancellation tests), severe functional disability (modified Rankin Scale score > 4), history of neurological disorders, confounding (Geriatric Depression Scale score < 11), pregnancy, or electrical implant devices such as pacemakers or defibrillators were also excluded from participation.

| Patient | Sex | Age | Stroke | Dominant Hand |
|---|---|---|---|---|
| 1 | M | 70 | Yes | Right |
| 2 | M | 71 | Yes | Right |
| 3 | F | 62 | Yes | Right |
| 4 | F | 75 | No | Right |
| 5 | M | 64 | No | Right |
| 6 | M | 53 | No | Right |

*Table 1:* Patients' Demographics

### Apparatus and Setup

We employed the Kinereach motion-tracking system[13] with integrated eye-tracking functionality for data collection. Participants were seated in a height-adjustable chair with their chin positioned and stabilized centrally. To minimize extraneous movement, all arm joints distal to the elbow were immobilized using an adjustable brace. Limb position and orientation were recorded at 116 Hz using four 6-degree-of-freedom magnetic sensors (trackSTAR; Ascension Technology) positioned on the hand and upper arms. Using digitized bony landmarks, we computed 10-degree-of-freedom arm movements, which enabled estimation of wrist, elbow, and shoulder joint positions.

To reduce friction and gravitational effects, each participant's hand rested on an air sled that provided continuous pressurized air, enabling friction-free movement. Task stimuli were displayed on an inverted HD monitor and projected onto a mirrored screen positioned at chin level, creating the visual illusion that elements appeared in the same horizontal plane as the hand. The participants' hands were occluded from direct view. A cursor represented the real-time position of the index finger, while start positions and target locations were indicated by circles that appeared at the beginning of each trial.

An EyeLink 1000 Plus eye tracker, mounted 52 cm from the participant's eyes and precisely angled to minimize occlusion, recorded eye movements. This configuration ensured that the eye tracker did not obstruct the projected stimuli, thereby preserving data accuracy when participants viewed the mirrored display.

The experimental system consisted of two synchronized computers: one dedicated to limb tracking and stimulus presentation, and another for eye tracking data acquisition. Custom MATLAB scripts (MathWorks Inc., Natick, MA, USA) and functions from the Psychophysics Toolbox managed eye-tracking stimulus presentation and data collection. Eye and limb tracking signals were synchronized through triggering mechanisms, ensuring precise temporal integration between the EyeLink and KineReach systems.

**Procedure**

**Calibration**

**Limb Tracking Calibration** Precise sensor placement was essential for accurate limb tracking. We positioned two sensors on each participant's limb: sensor 3 was attached to the right hand, sensor 4 to the upper right arm, and sensor 1 to the left hand. Sensor 2 served a dual purpose – it was initially used as a digitizing stylus to map anatomical landmarks, including finger joints, wrist, elbow, and shoulder positions. Following landmark digitization, sensor 2 was repositioned and secured to the left upper arm.

**Eye Tracking Calibration** Eye tracking calibration was performed using a customized protocol. Thirteen targets were presented in randomized order and displayed sequentially on the mirrored screen. Participants were instructed to fixate on each target as it appeared. Calibration accuracy was subsequently evaluated using custom scripts that calculate the difference between the recorded gaze point and the actual target coordinates in both X and Y positions.

**Task Design and Conditions**

Each participant completed four task conditions:
In each condition and trial, a central 'fixation' point first appeared at screen center. Subjects were required to fixate this position while simultaneously covering it with the index finger. After a delay the movement was cued. Feedback, both auditory and visual, was provided following each reach. Feedback showed the reach trajectory and endpoint relative to the target, as well as whether the target was touched.

1. Natural Reach: A fixation point appeared at screen center. After a delay, a peripheral target appeared. Participants were instructed to reach to the target by moving both eyes and hand as they normally would.
2. Segmented Look-Then-Reach: First the peripheral target appeared along with a saccade cue, and after a second delay a second auditory tone was played, cueing the reach. Participants were told they must wait to reach until hearing the second auditory cue.
3. Reach-Only: First the peripheral target appeared along with a saccade cue. Participants were instructed to make a reach to the target without moving their eyes away from the initial fixation position. All participants completed 32 trials per condition, evenly split between left and right targets. Trial order was pseudorandomized and conditions were blocked.

**Data Processing and Analysis**

Saccade onset was defined as the time point at which eye velocity exceeded an adaptive threshold[14]. Reach onset was defined as the time the index finger left the start zone (radius = 1 cm) and velocity exceeded 5 cm/s. For each trial, saccade and reach latencies were calculated relative to the corresponding auditory cuing of saccade and/or reach movements.

Distributions of latencies are plotted as histograms in Fig. 2. To assess whether the temporal distribution of saccades was structured (i.e., time-locked to reach), we compared each individual's saccade timing distribution in the natural task to a uniform distribution over the analysis window from 0 to 3s using a Kolmogorov–Smirnov (KS) test. The same analysis was performed on control participants to confirm the presence of a structured (non-uniform) distribution. A significance threshold of $p < 0.05$ was used.

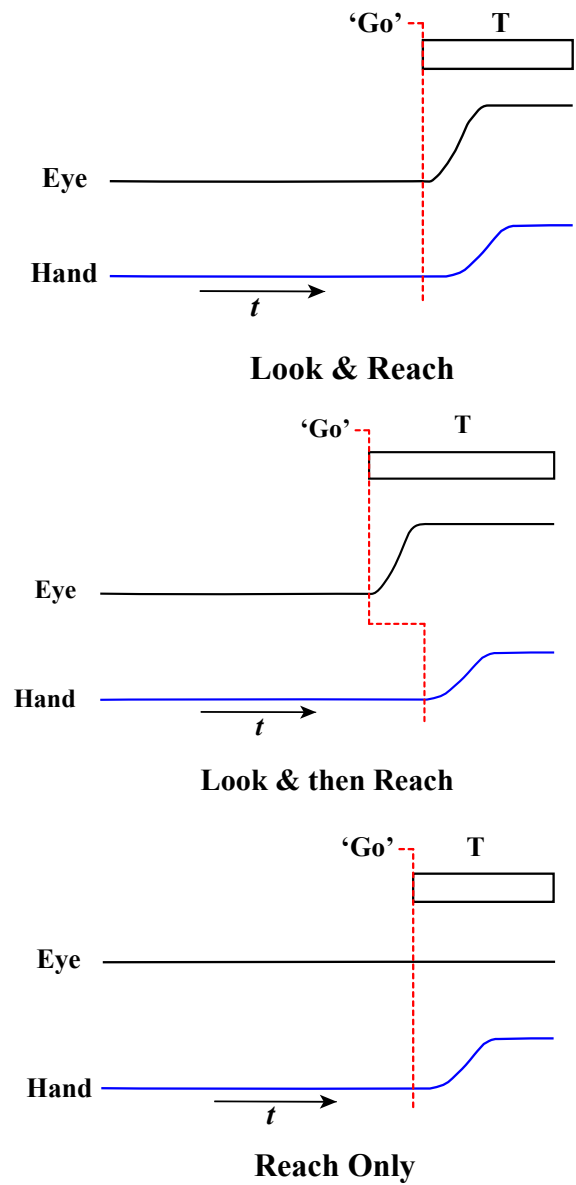

Figure 1:Fig. 1: (T) denotes the target onset; the Red dotted line represents the 'Go' Signal; (t) denotes time. The figure outlines four conditions: (i) Look and Reach (L&R),(ii) Look & then Reach (LtR), and (iii) Reach Only (RO) tasks.

**Results**
In the natural reach condition, control participants exhibited typical eye–hand coordination. Saccades reliably preceded reach onset by approximately 450 ms, producing a distribution with single peak centered near the time of reach initiation.

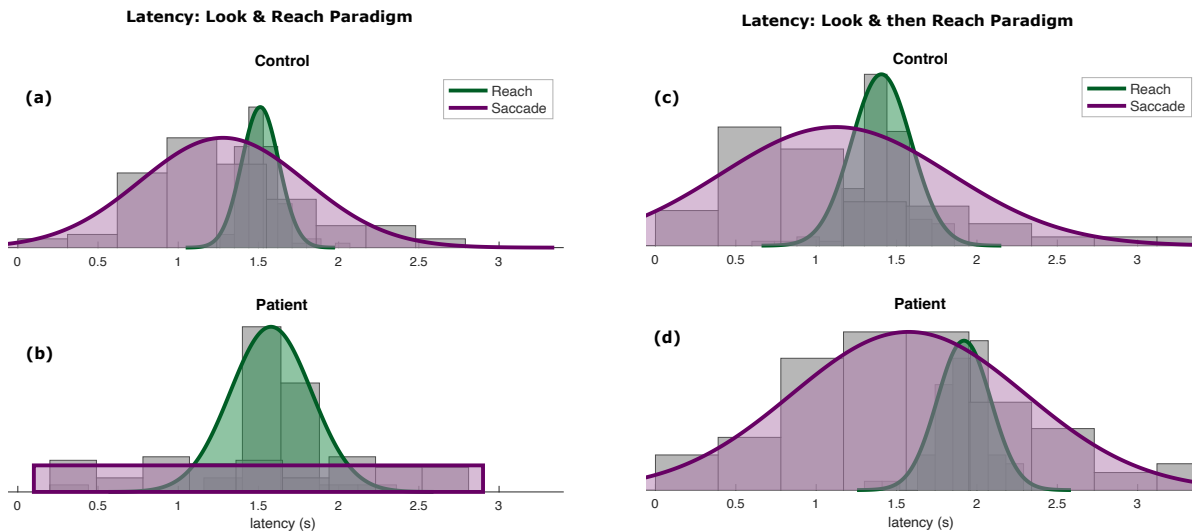

*Figure 3: Preliminary latency data. (a) Control reach and saccade latencies. (b) MCA reach and saccade latencies. (c), (d) Reach and saccade latencies in the segmented 'first-look and then-reach' task.*

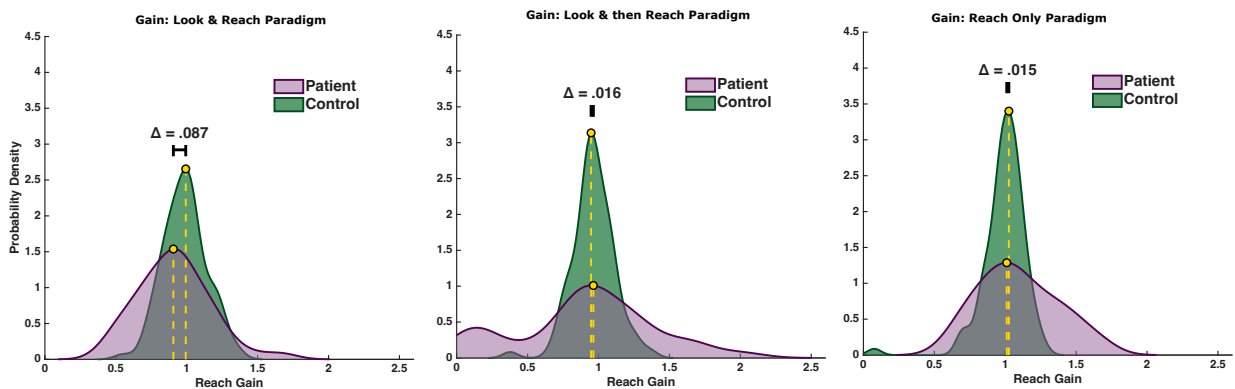

*Figure 2: Preliminary reach gain data. (a) 'Natural' simultaneous look & reach condition. (b) Segmented 'first-look, and then-reach' condition. (c) 'Reach-only' condition (gaze stable at reach start position).*

In contrast, stroke participants showed a flat distribution of saccade timing, with no clear peak or alignment to the onset of the reach, consistent with prior observations that eye and hand coordination is often disrupted following acquired brain injury[10]. Despite this temporal disorganization, reach timing in stroke participants remained singly peaked and similar in latency to that of controls (mean latency = 1.50s vs. 1.46s, n.s.), indicating preserved motor initiation but impaired inter-effector coordination. Statistical testing confirmed this dissociation. In stroke participants, the distribution of saccade times could not be distinguished from a uniform distribution (Kolmogorov–Smirnov test, $p > 0.2$), whereas in controls, the saccade distribution—and the reach distributions in both groups—differed significantly from uniform ($p < 0.001$), demonstrating a severe decoupling of the typical relationship between saccade and reach timing in the stroke group.

In terms of spatial performance, reach gain, the primary index of functional accuracy, revealed that stroke participants' reaches were significantly hypometric compared to those of controls ($t(18) = 2.5$, $p < 0.05$), consistent with mild impairment in motor execution.

Importantly, in the segmented look-then-reach condition, stroke participants demonstrated a restoration of a unimodal saccade timing distribution, time-locked to the saccade cue, consistent with previous findings that structured visual and auditory cues improve eye-hand synchrony[11]. Under these conditions, reach gains in stroke and control participants were statistically indistinguishable (t(18) = 1.0, p > 0.2), suggesting that both spatial and temporal components of eye-hand behavior were normalized when the task structure externally imposed temporal segmentation.

Also consistent with previous findings[15-17], the variance of reach timing improved in segmented relative to the natural condition for stroke participants and deteriorated for control participants, suggesting that disrupting normal inter-effector timing during a reach is detrimental within the intact nervous system, but improves performance when the neural substrates underlying inter-effector timing are themselves disrupted.

Together, these findings demonstrate that the ability to generate timely, goal-directed saccades is preserved after stroke, but the spontaneous temporal linkage between saccade and reach initiation is selectively disrupted. External structuring of movement timing can restore both coordination and performance accuracy.

**Discussion**
Our findings reveal a striking dissociation in individuals with stroke: initiation of motor execution is relatively preserved, but coordination between effectors—particularly the temporal coupling of eye and hand movements—is disrupted with downstream effects on performance. Stroke participants were able to produce accurate reaches and saccades when movements were performed in isolation, including as a cued sequence of separate eye and hand actions (LtR condition), indicating that core effector functions remained intact. However, when spontaneous coordination was required in the natural, uncued condition, stroke participants consistently failed to generate the time-locked pre-reach saccade, as observed in controls. Whereas control participants initiated saccades with a stereotypical temporal relationship to reach onset, stroke participants showed a flat, unstructured distribution of saccade onsets, indicating that saccade timing was no longer functionally linked to reach initiation.

This profound decoupling emerged despite the fact that the distribution of reach latencies in the stroke group remained singly-peaked and marched to those of controls. Critically, the timing of saccades in the stroke group lost their predictive, anticipatory function. Thus, the fundamental mechanism of coordination—particularly in naturalistic, self-governed tasks—is selectively impaired after stroke, despite intact individual movement capacity.

These findings align with a growing body of work suggesting that stroke recovery cannot be fully understood by assessing movement of individual effectors in isolation[18,16]. Effective coordination depends on the integration of spatial and temporal information within a distributed frontoparietal sensorimotor network, including the posterior parietal cortex, which is pivotal for mapping vision to action and for synchronizing effectors[4,5,19]. Stroke-related lesions in these regions—especially within the posterior parietal cortex—can thus lead to selective disruption of temporal eye–hand coupling even in the absence of overt limb weakness or oculomotor deficits[20].

Understanding such dissociations post-stroke has both theoretical and clinical significance. Theoretically, it suggests that control of timing between effectors is not an automatic consequence of effector integrity, but a distinct function that can be selectively impaired. Clinically, this finding highlights the need for rehabilitation approaches that go beyond strengthening individual effectors to restoring the natural coordination between them. In real-world tasks—such as picking up an object while navigating a cluttered space—failure to coordinate eye and hand movements efficiently may compromise safety and independence[11], even in patients with relatively mild motor impairment (patients who demonstrate largely intact limb strength and unimpaired oculomotor range). Identifying and addressing these subtle yet impactful deficits may be essential for developing new rehabilitation treatments and optimizing post-stroke recovery.

Assessment and Rehab

We found that saccade timing was normalized in the segmented, cue-driven condition, indicating that the underlying oculomotor system remains trainable and intact after stroke. In structured contexts, patients were able to generate appropriately timed saccades to targets. But in natural conditions, without external scaffolding, they failed to initiate gaze shifts timed to the upcoming reach. This pattern of preserved execution but impaired spontaneous coordination points to a loss of automatic integration, rather than a general impairment in movement generation. This has great implications when it comes to assessment and rehabilitation treatments.

Diagnostics

Assessment in stroke patients should place focus on coordination rather than motor function alone. Although patients had intact movement components, they did not have intact motor integration. Assessment protocols in stroke rehabilitation may benefit from incorporating measures of inter-effector timing, rather than focusing solely on endpoint performance. A patient who can generate fast, accurate reaches may still experience significant deficits in eye-hand coordination that go undetected in conventional assessments. These timing disruptions could underlie subtle impairments in real-world function or contribute to learned non-use of the affected limb in visually guided tasks. Additionally, in more demanding time-limited tasks in which foveation is critical such as when catching a ball, failure to precisely time the required saccade may lead to task failure or even injury[21,22].

Therapeutics

This dissociation between motor execution and coordination in stroke opens the door for new rehabilitation treatments. Such treatments should be aimed not only at motor strength and speed, but at training and re-establishing the natural temporal structure of action[23,24]. Particularly, structured cueing of saccade-reach sequences—such as with auditory prompts or visual highlighting—might serve to resynchronize decoupled systems and reinforce natural coordination patterns[15,16]. Over time, these structured interventions may promote re-engagement of internal coupling mechanisms, facilitating a transition back to automatic integration during natural behavior.

Functionally, impaired anticipatory gaze shifts have significant real-world consequences. Normally, fixating a target just prior to a reach supports both reach planning and online reach correction. In the latter case, target fixation brings the to-be-manipulated object within the high-acuity foveal region of the retina and appears to act as an attractor for the reach endpoint - both of which facilitate precise object interaction. In the absence of such coordination, stroke patients may engage in suboptimal or inefficient motor strategies, especially in dynamic or visually complex environments. Tasks such as reaching for moving objects, navigating crowded spaces, or using tools could be disproportionately affected, despite preserved limb strength and range of motion[2,3].

Limitations of this study include the modest sample size, lack of lesion mapping, and focus on chronic stroke. Future work should leverage neuroimaging (e.g., lesion-symptom mapping, structural connectivity) and study patients in the acute and subacute phases, as emerging evidence highlights sensitive windows of recovery and potential for intervention far beyond the traditional 'critical window' after stroke.

In conclusion, our findings highlight a dissociation between motor execution and coordination in stroke, emphasizing that intact movement components do not guarantee intact motor integration. By identifying a reversible, cue-sensitive disruption in saccade timing linked specifically to its coordination within a reaching movement, this work points toward new opportunities for rehabilitation aimed not only at motor strength and speed, but at training and re-establishing the natural temporal structure of action.